\documentclass[onecolumn]{elsart3p}

\usepackage{graphicx}
\usepackage{amssymb}
\usepackage{natbib}		
\usepackage{subfigure}	        
\usepackage{aas_macros}
\usepackage{multirow}

\begin{document}

\begin{frontmatter}

\title{Low energy polarization sensitivity of the Gas Pixel Detector}
\author[IASF,UTOV]{F.~Muleri\corauthref{cor}},
\corauth[cor]{Corresponding author.}
\ead{fabio.muleri@iasf-roma.inaf.it}
\author[IASF]{P.~Soffitta},
\author[INFN]{L.~Baldini},
\author[INFN]{R.~Bellazzini},
\author[INFN]{J.~Bregeon},
\author[INFN]{A.~Brez},
\author[IASF]{E.~Costa},
\author[IASF]{M.~Frutti},
\author[INFN]{L.~Latronico},
\author[INFN]{M.~Minuti},
\author[ASI]{M.~B.~Negri},
\author[INFN]{N.~Omodei},
\author[INFN]{M.~Pinchera},
\author[INFN]{M.~Pesce-Rollins},
\author[INFN]{M.~Razzano},
\author[IASF]{A.~Rubini}
\author[INFN]{C.~Sgr{\'o}},
\author[INFN]{G.~Spandre},
\address[IASF]{IASF/INAF, Via del Fosso del Cavaliere 100, I-00133 Roma, Italy}
\address[UTOV]{Universit\`{a} di Roma Tor Vergata, Dipartimento di Fisica, via della Ricerca Scientifica 1, I-00133 Roma, Italy}
\address[INFN]{INFN sez. Pisa, Largo B. Pontecorvo 3, I-56127 Pisa, Italy}
\address[ASI]{ASI, Agenzia Spaziale Italiana, Viale Liegi 26, I-00198 Roma, Italy}

\begin{abstract}
An X-ray photoelectric polarimeter based on the Gas Pixel Detector has been proposed to be included in many upcoming space missions to fill the gap of about 30~years from the first (and to date only) positive measurement of polarized X-ray emission from an astrophysical source. The estimated sensitivity of the current prototype peaks at an energy of about 3~keV, but the lack of readily available polarized sources in this energy range has prevented the measurement of detector polarimetric performances.

In this paper we present the measurement of the Gas Pixel Detector polarimetric sensitivity at energies of a few keV and the new, light, compact and transportable polarized source that was devised and built to this aim. Polarized photons are produced, from unpolarized radiation generated with an X-ray tube, by means of Bragg diffraction at nearly 45$^\circ$. The diffraction angle is constrained with two orthogonal capillary plates, which allow good collimation with limited size thanks to the 10~$\mu$m diameter holes. Polarized photons at energy as low as a few keV can be produced with a proper choice of diffracting crystal, while the maximum energy is limited by the  X-ray tube voltage, since all the orders defined by the crystal lattice spacing are diffracted. The best trade-off between reasonable fluxes and high degree of polarization can be achieved selecting the degree of collimation provided by capillary plates.

The employment of mosaic graphite and flat aluminum crystals allow the production of nearly completely polarized photons at 2.6, 3.7 and 5.2~keV from the diffraction of unpolarized continuum or line emission. The measured modulation factor of the Gas Pixel Detector at these energies is in good agreement with the estimates derived from a Monte Carlo software, which was up to now employed for driving the development of the instrument and for estimating its low energy sensitivity. In this paper we present the excellent polarimetric performance of the Gas Pixel Detector at energies where the peak sensitivity is expected. These measurements not only support our previous claims of high sensitivity but confirm the feasibility of astrophysical X-ray photoelectric polarimetry.
\end{abstract}


\end{frontmatter}


\section{Introduction}\label{sec:Introduction}

The study of the polarization of X-ray emission has been recognized as a fundamental tool in investigating the geometry and the physics of astrophysical sources \citep{Rees1975,Meszaros1988, Weisskopf2006}. The degree and the direction of polarization add two parameters to the current observable quantities thus helping in the discrimination between different models which can be otherwise equivalent on the basis of the spectroscopic and timing observations. However this branch of astronomy has suffered limitations due to the lack of sensitivity of the previously flown X-ray polarimeters. As a matter of fact, it has been possible to register the only positive detection of X-ray polarization in the emission of the Crab Nebula \citep{Weisskopf1976, Weisskopf1978}. The development of photoelectric polarimeters now provides for a valuable alternative to the classical measurement techniques, i.e Bragg diffraction and Thomson scattering. In particular, the Gas Pixel Detector \citep{Costa2001,Bellazzini2006, Bellazzini2006c} allows a large increase of sensitivity and it will enable the detection of the polarized emission in galactic and extragalactic bright sources in a few days at the focus of a small X-ray optics \citep{Costa2006}, while the measurement of faint extragalactic sources could be performed with a large size mission like XEUS \citep{Bellazzini2006b}.

The employment of this new generation instrument on-board the upcoming space missions requires for the first time the study of its response to polarized and unpolarized radiation in the whole energy range of interest. While the use of the Gas Pixel Detector is actually taken into account even at higher energies \cite{Muleri2006}, it is currently planned between $\sim$2 and $\sim$10~keV with a mixture of Helium and DME gas, in order to be able to exploit the classical X-ray optics band-pass and the higher flux from astrophysical sources at low energy. The development of the Gas Pixel Detector hence demands a polarized source in this energy range.

While synchrotron facilities can readily produce polarized radiation at energies of a few keV, they can hardly be employed for the continuous and time-consuming development of a new generation detector as the Gas Pixel Detector. The requirement of a compact and transportable polarized source was first attained with a source based on Thomson scattering \citep{Costa1990}. Unpolarized photons, produced with an X-ray tube, are scattered on a lithium target, enclosed in beryllium to prevent its oxidation and nitridation, and the geometry of the source constrains the scattering angle at nearly 90$^\circ$. Since photons scattered at 90$^\circ$ result completely polarized in the direction orthogonal to the scattering plane, the output radiation is highly polarized.

The main issue behind this kind of technique is that scattered radiation is not monochromatic and, above all, its employment is limited at low energy by the photoelectric absorption in the lithium target. A Thomson-scattering source has been successfully employed to measure the polarimetric sensitivity of the Gas Pixel Detector with a chromium (5.41~keV) and copper (8.04~keV) X-ray tube, while a Monte Carlo software was used to extend the measured performances at lower energies. According to the Monte Carlo results, the maximum sensitivity is reached at about 3 keV, and we expect that polarimetric response is rather variable at these energies because of the fast increase of the modulation factor. Hence its measurement at energies of a few keV is of fundamental importance to characterize the polarimetric sensitivity of the Gas Pixel Detector. Moreover many authors have stressed the importance of spectro-polarimetry in the study of astrophysical sources \citep{Rees1975,Connors1980,Mitrofanov2003} which requires a precise knowledge of the instrumental response with energy.

To meet this requirement, we devised and built a new polarized source, based on the Bragg diffraction at 45$^\circ$. The proper choice of the diffracting crystal allows the production of low energy polarized photons. We employed mosaic graphite and flat aluminum crystals to generate 2.6, 3.7 and 5.2 keV radiation, which results almost completely polarized, and it allowed the characterization of the Gas Pixel Detector in the energy range where maximum sensitivity and major use is expected.

In Sec.~\ref{sec:ExpectedPerformance} we summarize the sensitivity of the Gas Pixel Detector as obtained from Monte Carlo software, while in Sec.~\ref{sec:BraggSource} we describe the Bragg polarized source. In Sec.~\ref{sec:Result} we report the polarimetric response of the Gas Pixel Detector at low energy and in Sec.~\ref{sec:Conclusion} we draw our conclusions.

\section{Expected low energy polarimetric sensitivity of the Gas Pixel Detector} \label{sec:ExpectedPerformance}

\subsection{The Minimum Detectable Polarization}

X-ray polarimeters generally measure the direction and the degree of polarization of incident radiation by means of the phase and the amplitude of their modulated response. The sensitivity of this kind of instrument is commonly expressed with their \emph{Minimum Detectable Polarization} (MDP), which represents the amplitude of the response modulation for unpolarized incident radiation only due to statistical fluctuations. Hence, only the detection of a modulation greater than MDP is statistically significant.

Assuming Poisson statistics and 99\% confidence level, we have:
\begin{equation}
MDP =\frac{4.29}{\epsilon \mu F} \sqrt{\frac{B + \epsilon F}{S T}}
\end{equation}
where $F$ and $B$ are respectively the source and the background flux in a selected energy range, $\epsilon$ the detector efficiency, $S$ the collecting area, $T$ the observing time and $\mu$ the modulation factor, i.e. the amplitude of the response modulation for completely polarized incident radiation. 

The modulation factor is a key-parameter for the sensitivity of a polarimeter. A larger value of the modulation factor implies lower statistical fluctuations and hence the capability to detect lower polarized signals.

Since the Gas Pixel Detector will be placed at the focus of an X-ray optics, the background, $B$, will be negligible and hence the MDP results inversely proportional to the so-called \emph{quality factor} $\mu \sqrt{\epsilon}$. The characteristics of the Gas Pixel Detector are to be chosen such that the quality factor within a selected energy range is maximized while at the same time in tune with the optical band-pass. 

\subsection{The modulation and the quality factor of the Gas Pixel Detector}

\begin{figure}[htbp]
\begin{center}
\includegraphics[angle=0,width=10cm]{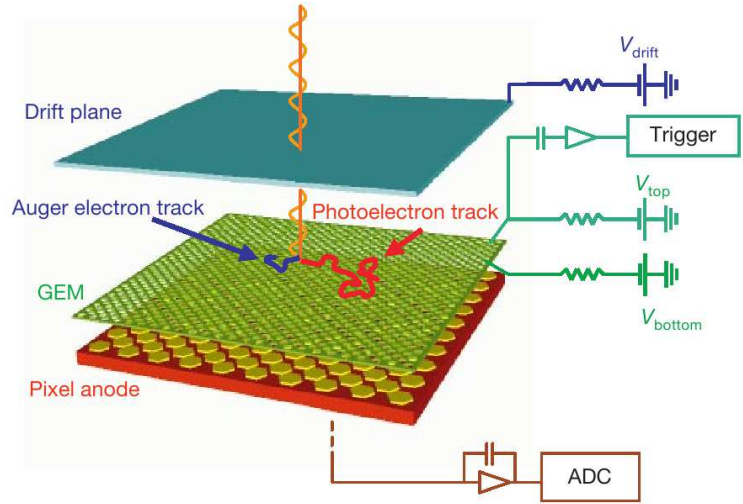}
\end{center}
\caption{\small Principle of operation of the Gas Pixel Detector. When a photon is absorbed in the gas, a photoelectron is emitted. As the latter propagates, it  is scattered by charges in the nuclei and loses its energy by ionization. The generated electron-ion pairs, which trace the photoelectron path, are amplified by the Gas Electron Multiplier (GEM) and collected on a fine sub-divided pixel detector. The polarization is derived from the reconstruction of the direction of emission of the photoelectrons, that carry a significant memory of the absorbed photons polarization. The ejection of the photoelectron is very likely followed by the production of an Auger electron, which must be distinguished because its direction of emission is isotropic. \label{fig:GPD}}
\end{figure}

The operating principle of the Gas Pixel Detector is based on the photoelectric absorption in a gas \citep{Costa2001}. The photoelectrons are preferencially emitted along the electric field of the absorbed photon, i.e. in the direction of polarization. Hence, when the incident radiation is linearly polarized, the emission of the photoelectrons is modulated with an amplitude proportional to the degree of polarization. The Gas Pixel Detector measures the amplitude and the phase of this modulation through the electron-ion pairs produced in the gas by photoelectron ionization. The pairs are amplified by an electric field generated with a Gas Electron Multiplier and collected on a fine sub-divided pixel detector (see Fig.~\ref{fig:GPD}). Since photoelectrons are subject to nuclear scatterings in the gas, the reconstruction of their direction of emission with an automatic algorithm is effective only if the tracks are resolved by the detector on many pixels. The Gas Pixel Detector, developed by the INFN of Pisa, was the first device that was capable of resolving photoelectron tracks on tens of pixels, even at low energy, thanks to its 50~$\mu$m pixel size.

The amplitude of the modulation measured by the instrument when completely polarized photons are absorbed, also known as the modulation factor, increases as the resolution and reconstruction of the photoelectrons tracks improves.
High energy photons produce longer photoelectrons tracks, therefore the modulation factor increases with photon energy yet it is inversely correlated with the efficiency.
In order to improve the modulation factor it is important to reduce the nuclear scattering of the photoelectrons in the gas and thus use mixtures with low atomic number components while on the other hand an increase in the atomic number improves the efficiency. Hence, a trade-off between modulation factor and efficiency is required to maximize the quality factor.

The quality factor has been calculated for a wide class of different gas mixtures with a Monte Carlo software \citep{Pacciani2003, Bellazzini2003}, which takes into account photoelectric absorption, photoelectron propagation and diffusion in the gas. The modulation factor at 5.41~keV, and occasionally at 8.04~keV, has been measured for the current 105k pixel 50~$\mu$m pitch detector filled with a subset of the most efficient mixtures \citep{Bellazzini2006}. Polarized photons were generated from 90$^\circ$ Thomson scattering of unpolarized radiation, produced with chromium or copper X-ray tube, on a lithium target. This source was inefficient at lower energies because of the overriding contribution of photoelectric absorption with respect to Thomson scattering on the target. Nevertheless the measured value of the modulation factor, $\mu$, supported Monte Carlo results at these energies. We assumed so far that this validation could be extended to lower energies and made an intense use of the simulations as a tool for the optimization of the detector and for the proposal of various polarimeters on-board space missions. But, as outlined by \cite{Weisskopf2006}, a direct measurement in the range of major use was still missing.

\begin{figure}[htbp]
\begin{center}
\subfigure[\label{fig:EffMu}]{\includegraphics[angle=90,totalheight=5.7cm]{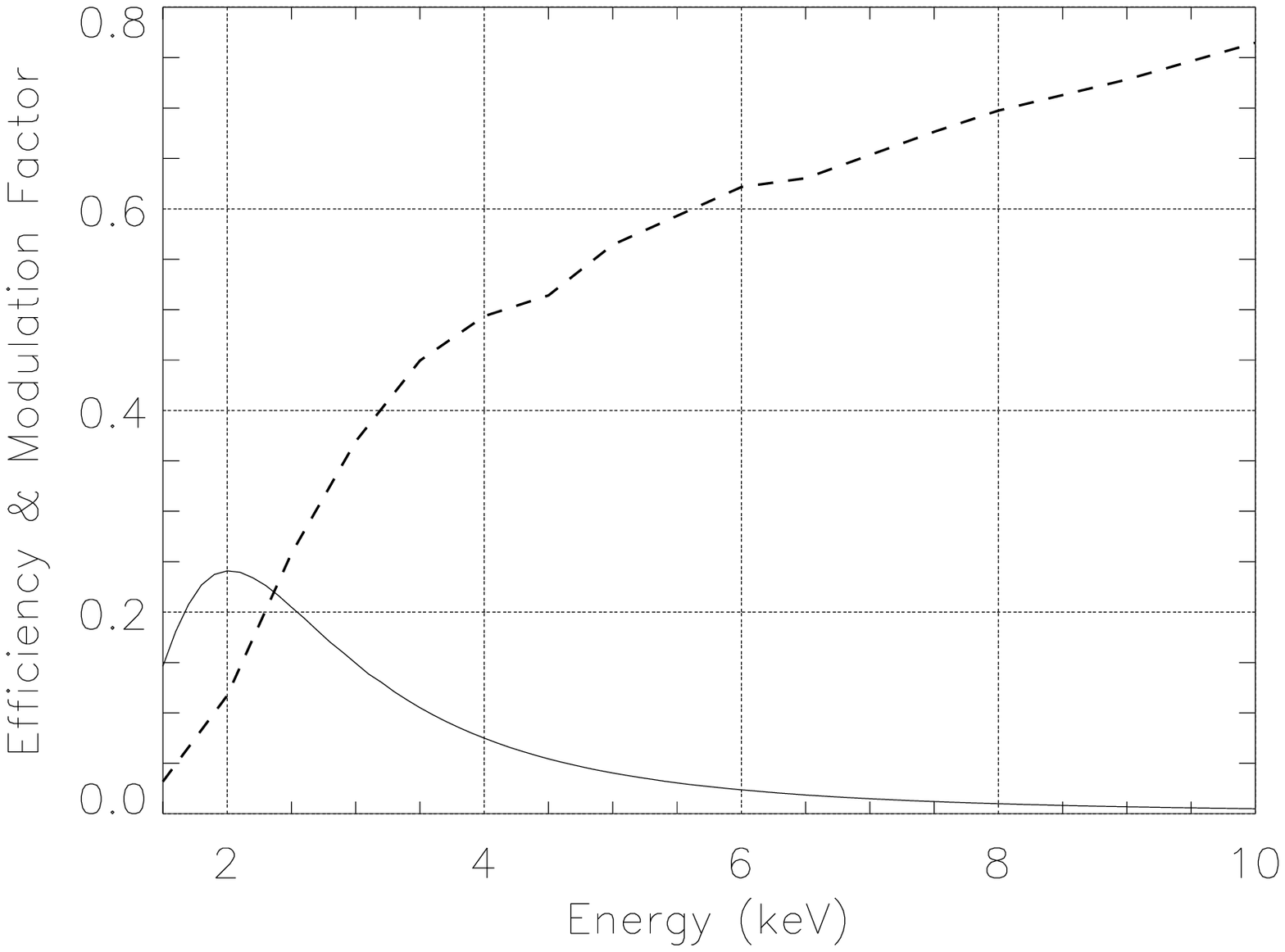}}
\subfigure[\label{fig:QualityFactor}]{\includegraphics[angle=90,totalheight=5.7cm]{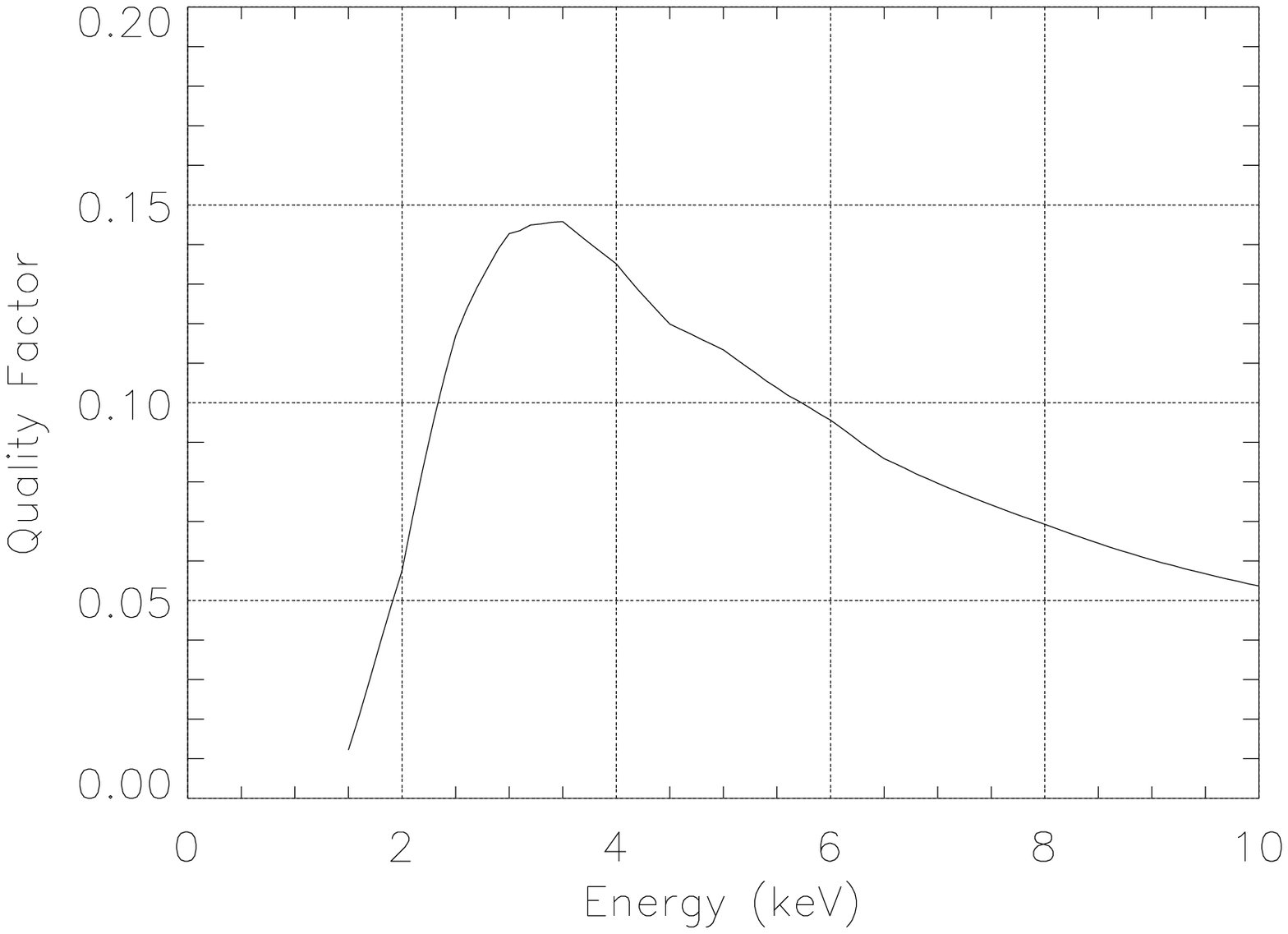}}
\end{center}
\caption{\small ({\bf a}) Efficiency (solid line), including 50~$\mu$m beryllium window, and modulation factor (dashed line) for the Gas Pixel Detector filled with 20\% He and 80\% DME mixture. Modulation factor is calculated with a Monte Carlo software. ({\bf b}) Quality factor for 20\% He and  80\% DME. The peak of the sensitivity is reached at about 3 keV.}
\end{figure}

The best trade-off in the energy range 2-10 keV, where an X-ray optics can be easily exploited, is obtained with a mixture of 20\% Helium and 80\% DME. The efficiency and the modulation factor, calculated with the Monte Carlo software, are plotted in Fig.~\ref{fig:EffMu}. The modulation increases quickly with energy and the quality factor, reported in Fig.~\ref{fig:QualityFactor}, peaks at about 3~keV.

The lack of an efficient and comfortable polarized source at energies lower than 5.4~keV prevented the measurement of the detector performances where maximum sensitivity is expected and the check of modulation increase with energy. This limitation is finally overcome with the source presented in Sec.~\ref{sec:BraggSource}.

\section{The Bragg diffraction polarized source} \label{sec:BraggSource}

\subsection{The Bragg diffraction at 45$^\circ$}

Bragg diffraction involves the diffraction of photons on the planes of a crystal lattice. If monochromatic radiation of unit intensity and energy $E$ is incident at a glancing angle $\theta$ to the diffraction planes of a flat crystal, it is diffracted at an angle $\theta$ and its intensity $P_E(\theta)$, called rocking curve, is strongly peaked at the angle which satisfies the Bragg condition \cite{Guinier1994}, relating the angle and the energy of the diffracted photons:
\begin{equation}
E = \frac{nhc}{2d\sin\theta}, \label{eq:BraggLaw}
\end{equation}
where $h$ and $c$ are respectively Planck's constant and the speed of light, $d$ the crystal lattice spacing and $n$ the diffraction order.

The efficiency of diffraction, i.e. the total intensity of diffracted radiation, for unpolarized and monochromatic radiation is a characteristic of the crystal only and it is commonly expressed with the integrated reflectivity $R_E$:
\begin{equation}
R_E = \int_0^\frac{\pi}{2} P_E \left(\vartheta\right) d\vartheta,
\end{equation}
where $\vartheta$ is the angle of incidence.

In Table~\ref{tab:IntegratedReflectivity} we report the theoretical integrated reflectivity for some crystals at 45$^\circ$ Bragg energy. In practice, the best integrated reflectivity is obtained with graphite crystal and it is close to its theoretical value listed in the Table.

\begin{table}[htbp]
\begin{center}
\begin{tabular}{r|c|c} 
Crystal                         & Energy$_{45^\circ}$                & $R_E \times 10^4$       \\
                                    & (keV)                                        &  (radians)                     \\
\hline
\hline
WS$_2$                       & 1.40                                          & 15.0                              \\
MoS$_2$                     &  1.43                                          &  8.3                               \\
PET                             &  2.01                                         &  4.3                                \\
Graphite (mosaic)       &  2.61                                         &  15.9                          \\
Calcite                         &  2.89                                          &  3.5                               \\
LiH                              &  4.3                                             &  35.0                            \\
LiF                               & 4.35                                           &  5.2                              \\
\end{tabular}
\end{center}
\caption{Theoretical integrated reflectivity for some crystals. It is assumed that the energy of the incident unpolarized photons is the Bragg energy for 45$^\circ$ diffraction. The value obtained in practice is close to that listed only for graphite and molybdenum disulfide (data from \cite{Silver1989}).} \label{tab:IntegratedReflectivity}
\end{table} 

\begin{figure}[tbp]
\begin{center}
\includegraphics[angle=0,width=12cm]{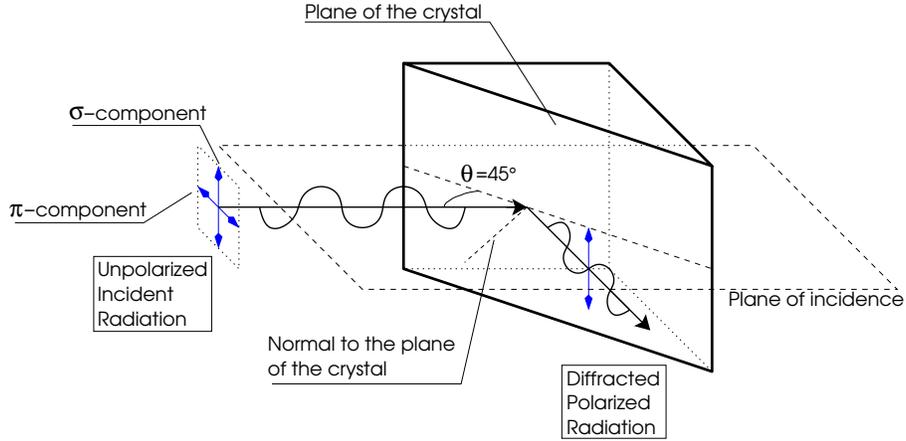}
\end{center}
\caption{\small Geometry of Bragg diffraction at 45$^\circ$. Unpolarized radiation is incident at 45$^\circ$ on the plane of a crystal. Since only the $\sigma$-component, orthogonally polarized to the plane of incidence defined by the direction of incidence and by the normal to the plane of the crystal, is efficiently diffracted, the output radiation is linearly polarized \label{fig:DP}}
\end{figure}

The integrated reflectivity depends on the direction of polarization of the incident radiation. Without loss of generality, electromagnetic waves can be considered as the superimposition of two waves orthogonally polarized. Hence we can decompose the incident radiation in two components, parallel ($\pi$-component) and perpendicularly ($\sigma$-component) polarized to the plane of incidence, which is the plane where the direction of the incident radiation and the normal of diffraction planes lie (see Fig.~\ref{fig:DP}). The ratio $k={R_E^\pi}/{R_E^\sigma}$ of the integrated reflectivity of $\pi$ and $\sigma$ components depends on the diffraction angle and, since it is generally lower than 1, the diffracted radiation is partially polarized perpendicularly to the plane of incidence. The degree of polarization is \cite{Evans1977}:
\begin{equation}
{\cal P} = \frac{1-k}{1+k}. \label{eq:PolarizationDegree}
\end{equation}

\begin{figure}[htbp]
\begin{center}
\begin{tabular}{c}
\subfigure[\label{fig:k}]{\includegraphics[angle=0,totalheight=6cm]{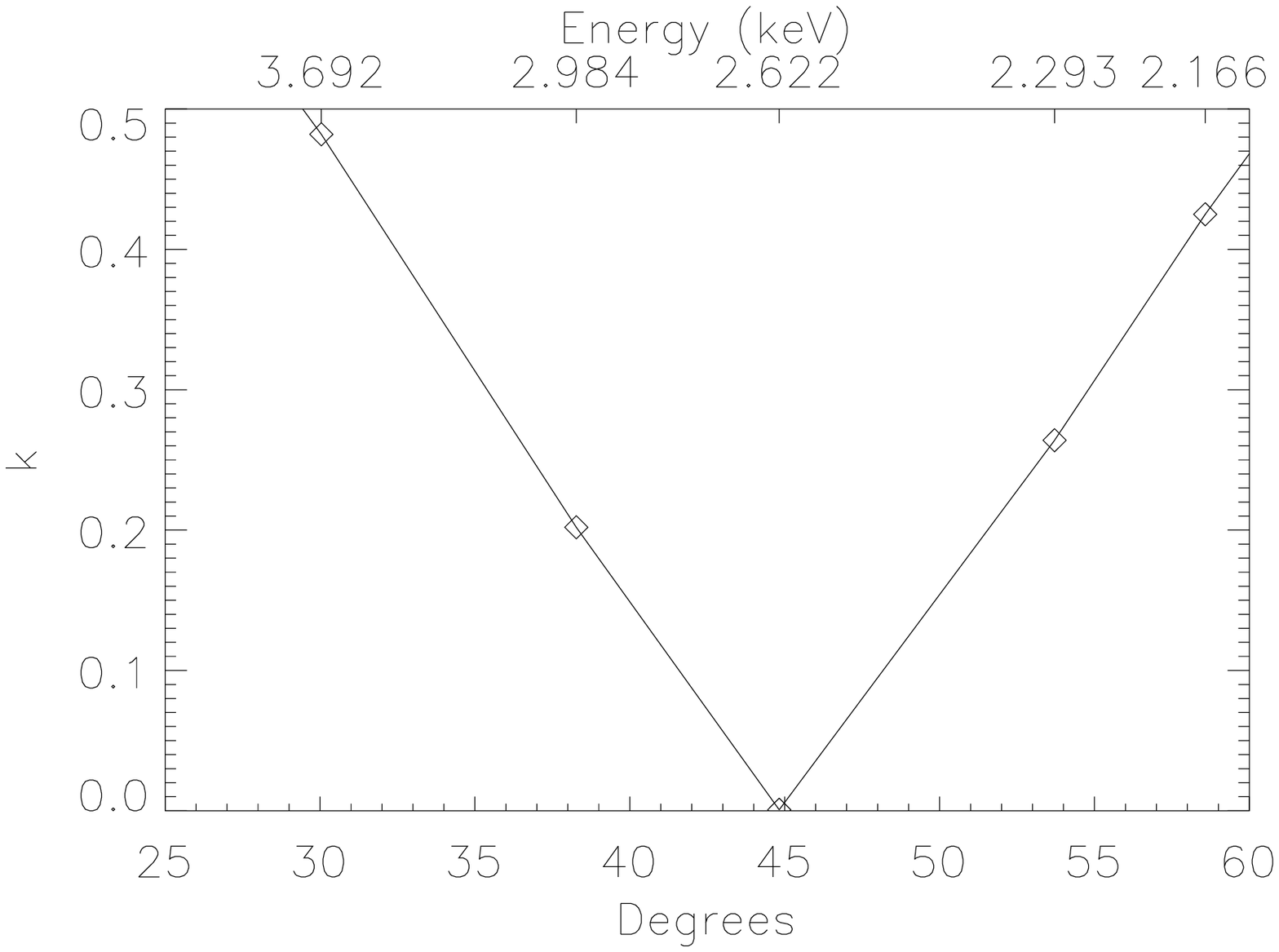}}
\subfigure[\label{fig:P}]{\includegraphics[angle=0,totalheight=6cm]{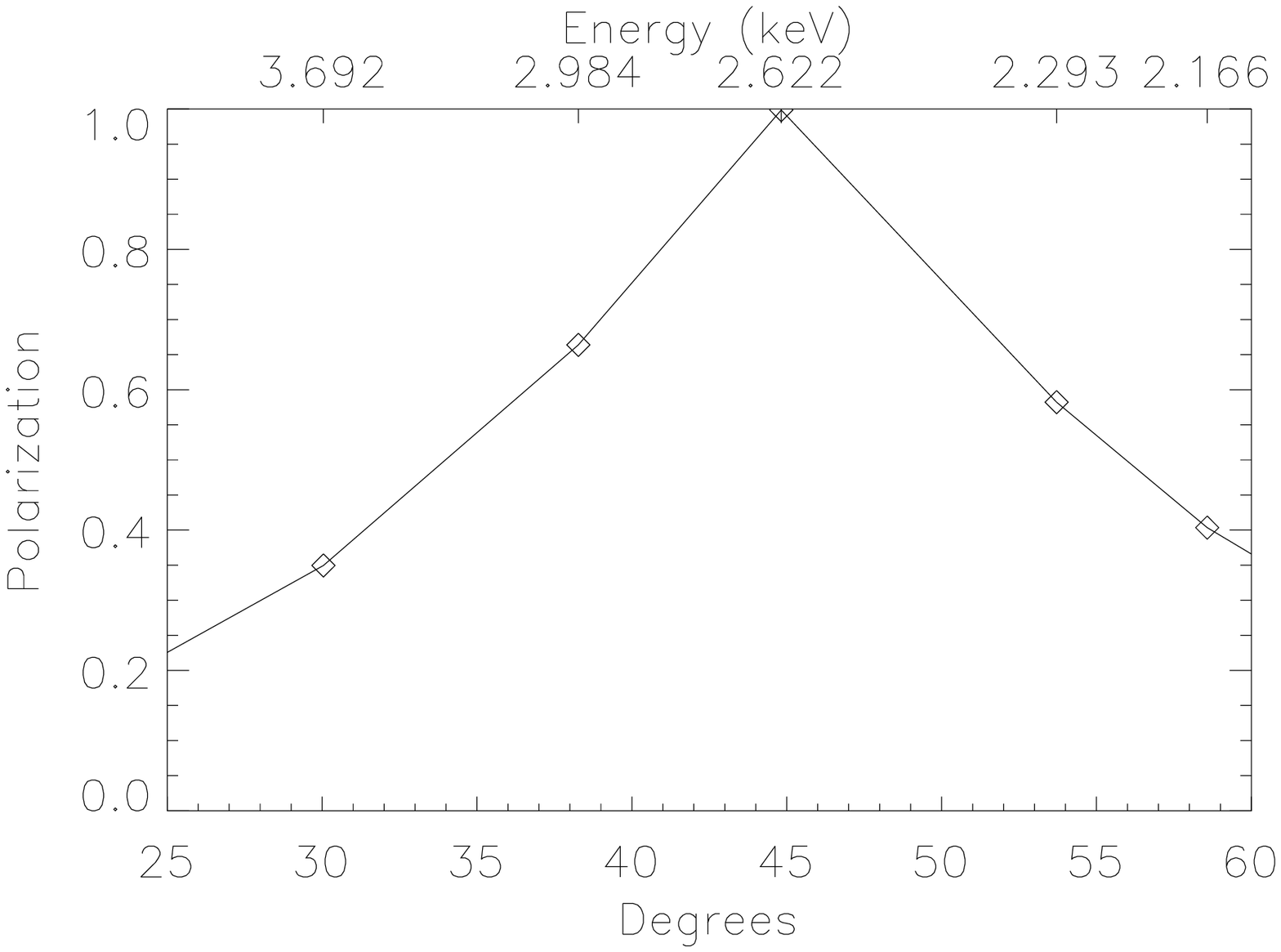}}
\end{tabular}
\end{center}
\caption{\small Production of polarized radiation by means of diffraction on a graphite crystals. ({\bf a}) Dependence of the ratio of integrated reflectivity for incident radiation completely polarized parallel and perpendicular to the plane of incidence with diffraction angle. ({\bf b}) Expected degree of polarization, derived from Eq.~\ref{eq:PolarizationDegree}. When the diffraction angle is 45$^\circ$, $k=0$ and the radiation is completely polarized. Diffraction angle and energy are related by Bragg's law (Eq.~\ref{eq:BraggLaw}). The value of $k$ has been calculated by \cite{Henke1993}.}
\end{figure}

In Fig.~\ref{fig:k} and Fig.~\ref{fig:P} we report the value of $k$, calculated by \cite{Henke1993}, and the value of the expected polarization in function of the diffraction angle $\theta$ for a graphite crystal. When the diffraction angle is 45$^\circ$, $k=0$ and the radiation is completely polarized orthogonally to the plane of incidence. Hence, constraining the diffraction angle to nearly 45$^\circ$, the radiation is almost completely polarized. Moreover since photons which are diffracted must satisfy, within some eV for flat crystals, the Bragg condition (Eq.~\ref{eq:BraggLaw}) for $\theta=45^\circ$ and an integer $n$, the spectrum is composed of equally spaced and nearly monochromatic lines.

We note that the large dependence of the value of $k$ with respect to $\theta$ requires that the constrain on the diffraction angle must be very tight in order to avoid the reduction of the mean polarization of the output radiation. This implies a very low value of the integrated reflectivity, with an energy band-pass of a few eV for a flat crystal, and hence a very low diffracted intensity for the incidence of continuum photons. In fact this was one of the major limitations for the X-ray polarimeters whose functionality was based on Bragg diffraction and were developed for the measurement of astrophysical polarized emission, constituted mainly by continuous spectra or broad cyclotron lines.

Integrated reflectivity can be increased, at the expense of a reduction of the degree of polarization, with the employment of mosaic crystals. In this case the crystal is not flat, i.e. formed by a plane parallel lattice, but it is composed of small domains each acting as an independent crystal. Domains are slightly and regularly misaligned, with a FWHM gaussian misalignment of about 1$^\circ$. Since the domains are smaller than the absorption length of X-ray photons, radiation crosses many domains before absorption and hence the diffraction can occur at an angle slightly different from 45$^\circ$, increasing the energy band-pass up to some tens of eV. In this case the energy width of the diffracted photons can be employed to estimate the mean degree of polarization of the output radiation.

High diffracted intensity and also high polarization can be achieved by the employment of flat crystals and monochromatic incident photons tuned with the Bragg energy at 45$^\circ$. In this case about half of the incident unpolarized radiation is diffracted. Moreover, knowledge of the photon energy allows to derive the diffraction angle and hence the degree of polarization from Bragg's law (Eq.~\ref{eq:BraggLaw}) and Eq.~\ref{eq:PolarizationDegree} respectively.

A convenient choice of the crystal allows the production of polarized radiation even at low energy. In Table~\ref{tab:Crystals} we report a list of fluorescence lines and crystal pairs which can be employed for the production of polarized photons at a few keV. The diffraction angle and the polarization, calculated from the well known fluorescence energy and from Eq.~\ref{eq:PolarizationDegree}, is also reported.

\begin{table}[htbp]
\begin{center}
\begin{tabular}{c|c|c|c|c} 
Fluorescence line                     & Energy (keV) & Crystal                              & $\theta$          & ${\mathcal P}$    \\
\hline
\hline
L$\alpha$ Molybdenum            & 2.293               & Rhodium (001)                 & 45.36$^\circ$  & 0.9994                 \\
K$\alpha$ Chlorine                   & 2.622               & Graphite (002)                  & 44.82$^\circ$   & 0.9986                 \\
L$\alpha$ Rhodium                  & 2.691               & Germanium (111)             & 44.86$^\circ$  & 0.9926                 \\
K$\alpha$ Calcium                    & 3.692               & Aluminum (111)              & 45.88$^\circ$   & 0.9938                 \\
K$\alpha$ Titanium                  & 4.511               & Fluorite CaF$_2$ (220)    & 45.37$^\circ$  & 0.9994                 \\
\end{tabular}
\caption{Tuning between fluorescence lines and diffracting crystals. $\theta$ is the diffraction angle and ${\mathcal P}$ the polarization of diffracted photons. Data from calculation performed by \cite{Henke1993}.} \label{tab:Crystals}
\end{center}
\end{table}

\subsection{The source design}


As was stated previously, the choice of the diffracting crystal as well as the choice of the incoming and output collimators allow not only to produce polarized photons at different energies but also a reasonable trade-off between the degree of polarization and the flux. It is exactly with this concept in mind that we designed a modular source thus allowing to easily change each component and fine tune the characteristics of the source to our needs. 

\begin{figure}[htbp]
\begin{center}
\subfigure[]{\includegraphics[angle=0, totalheight=6cm]{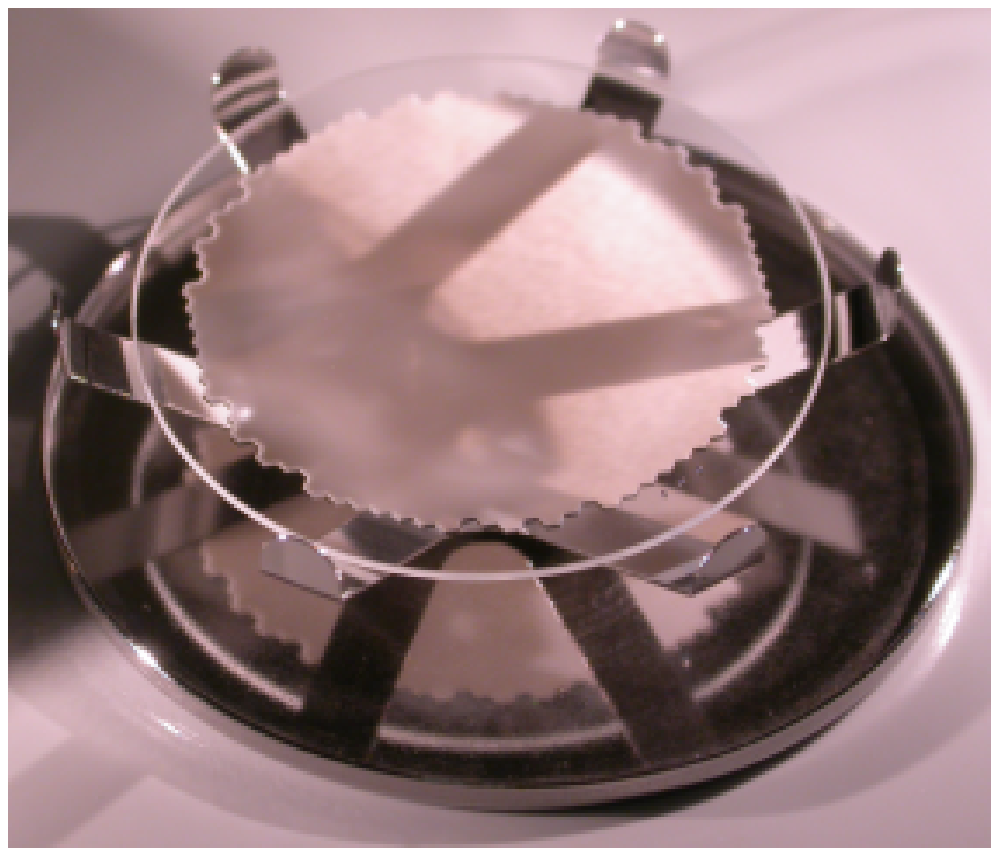}}
\subfigure[]{\includegraphics[angle=0, totalheight=6cm]{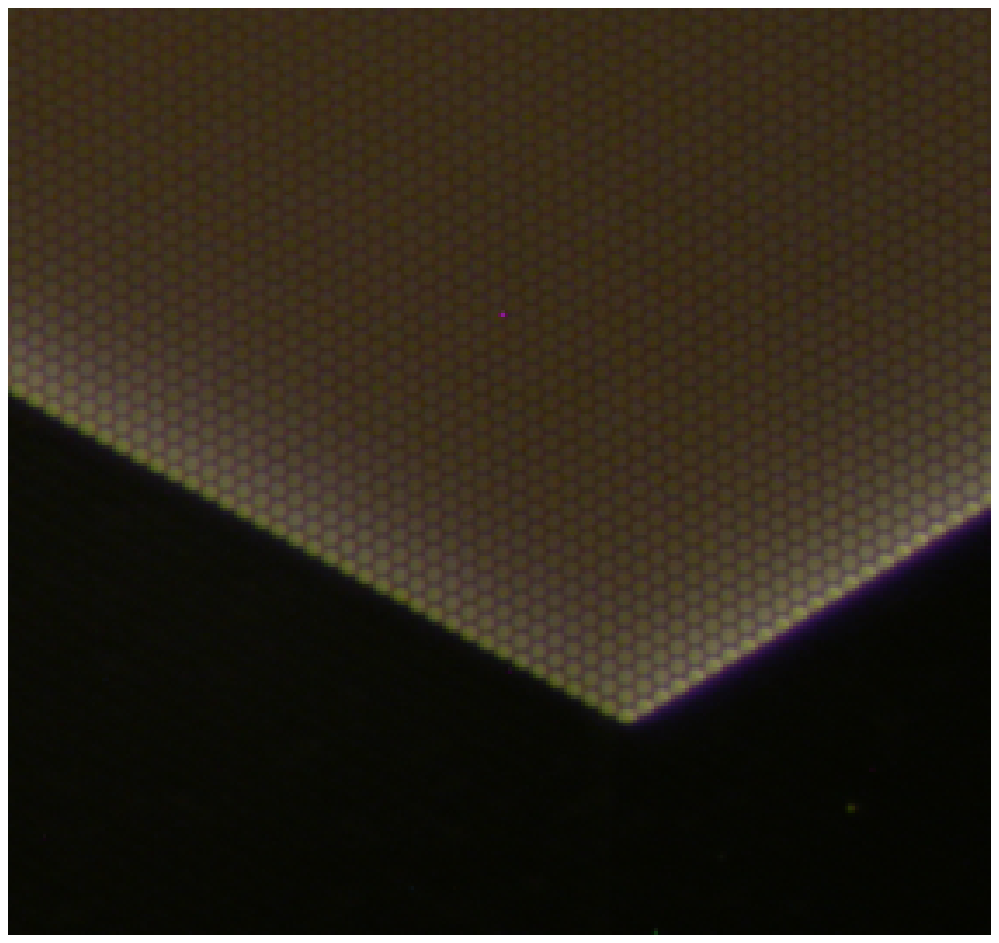}}
\end{center}
\caption{\small ({\bf a}) A capillary plate, built by Hamamatsu Photonics, employed as a collimator. ({\bf b)} An enlargement view showing the hexagonal pattern and the 10~$\mu$m holes. }\label{fig:Capillary}
\end{figure}

We collimated incoming and output X-rays with lead-glass capillary plates (see Fig.~\ref{fig:Capillary}), which allow good collimation with limited size. The 10~$\mu$m diameter holes, organized in a hexagonal pattern, allow an on-axis transparency of 57\%. We employed two types of capillary plates with 0.4 and 1~mm thickness, which narrow the outgoing beam with a semi-aperture of $\frac{1}{40}=1.4^\circ$ and $\frac{1}{100}=0.6^\circ$ respectively. The effective diameter of the capillary plates is about 20 and 27~mm for the broad and narrow capillary plate respectively, and hence can be used to completely cover the Gas Pixel Detector $15\times15~\mbox{mm}^2$ surface.

We employed a small and low power commercial (200 mW) X-ray tube with a calcium or copper anode to assure a good portability of the source. As reported in Table~\ref{tab:Crystals}, the calcium fluorescence K$\alpha$ line at 3.69~keV is well tuned with the 45$^\circ$ diffraction on the aluminum flat crystal and hence can be exploited for the efficient production of highly polarized photons at this energy. Being that there is a very narrow energy bandwidth for diffraction from a flat crystal, the contribution of the diffracted continuum from bremsstrahlung radiation\footnote{The spectrum of an X-ray tube is composed by lines, corresponding to electronic transition (fluorescence) of atoms in the target, and a continuum bremsstrahlung emission due to the electrons breaking in the target.} can be neglected with respect to the much more intense fluorescence emission. This assumption allows for the precise determination of the degree of polarization from Eq.~\ref{eq:PolarizationDegree}, which results in this case equal to 99.38\%.

The continuum from the bremsstrahlung emission from the copper tube was used for the diffraction on a mosaic graphite crystal, which allows the production of polarized photons at 2.6~keV (I order) and 5.2~keV (II order). Since the bremsstrahlung radiation increases with the square of the target atomic number, the employment of the copper anode tube is preferable with respect to the calcium one. To achieve a reasonable diffracted flux, we employed grade~B and D mosaic graphite crystals, with a misalignment FWHM of 0.8 and 1.2 degrees respectively, with both narrow and broad capillary plates.

\begin{figure}[htbp]
\begin{center}
\includegraphics[angle=0, width=7.5cm]{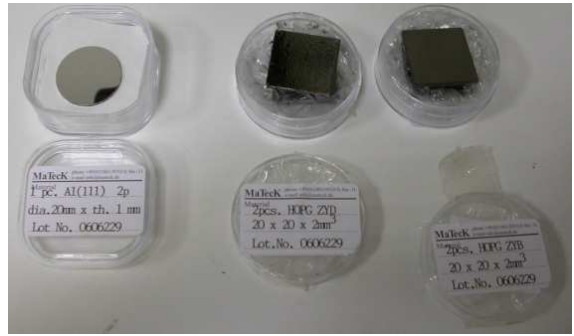}
\end{center}
\caption{\small From left to right: aluminum and graphite crystals built by Matek. Graphite crystals differ from the amplitude of domains misalignment.}\label{fig:Crystals}
\end{figure}

In Fig.~\ref{fig:Crystals} the aluminum crystal and the grade~B and D mosaic crystals are shown. The first is 20~mm in diameter, while the graphite crystals are $20\times20~\mbox{mm}^2$

\subsection{The source construction}

\begin{figure}[htbp]
\begin{center}
\subfigure[\label{fig:PolarizerTop}]{\includegraphics[angle=0,totalheight=7cm]{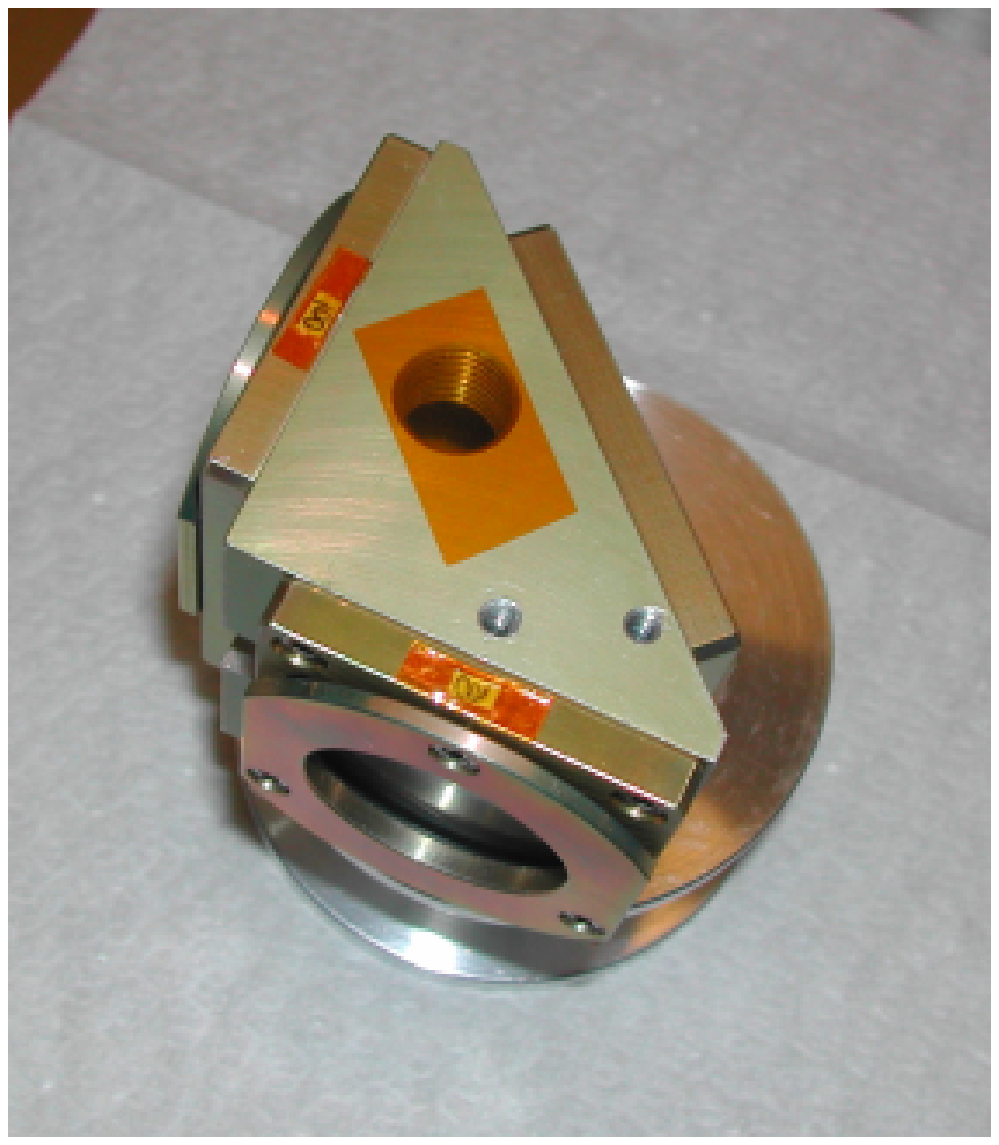}}
\subfigure[\label{fig:PolarizerSide}]{\includegraphics[angle=0,totalheight=7cm]{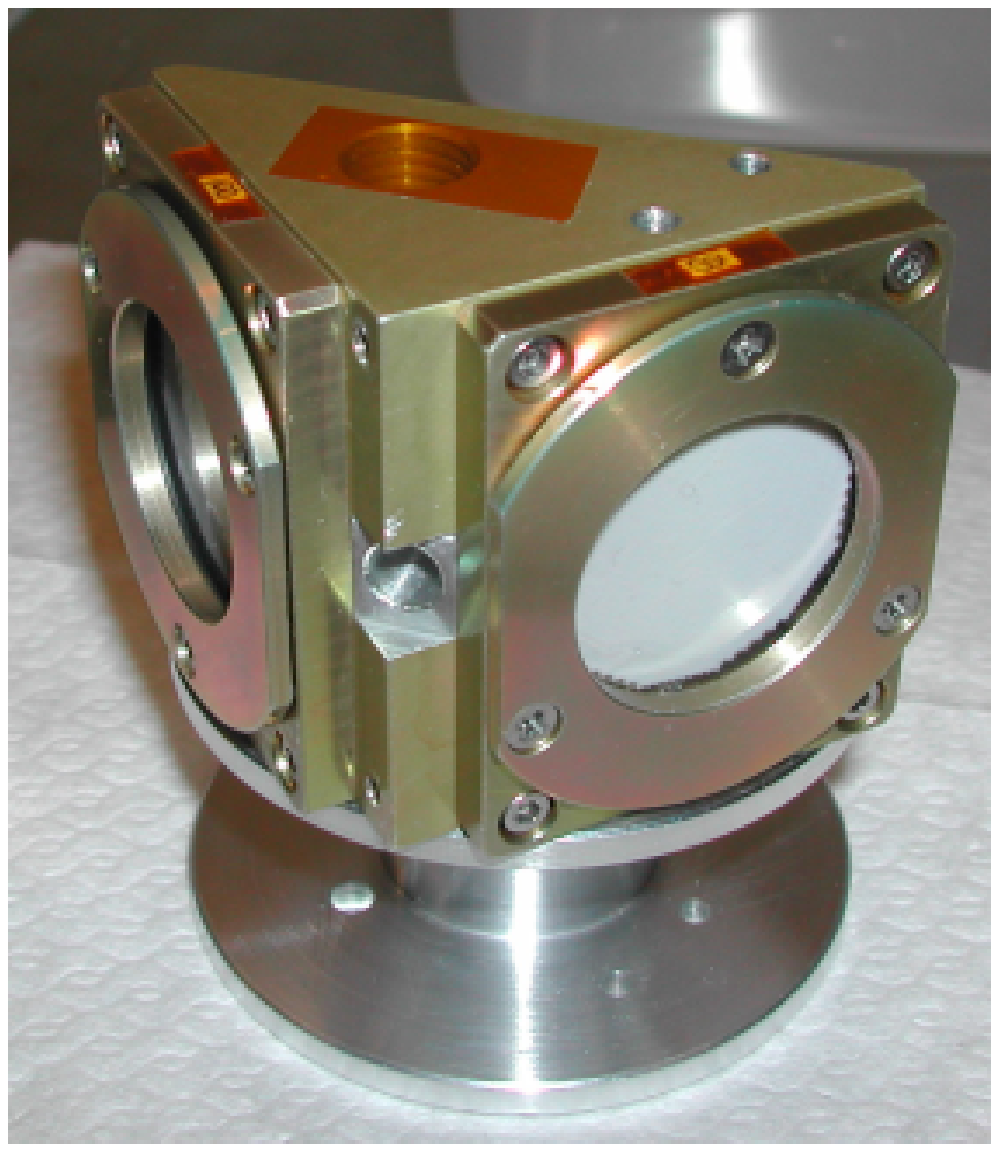}}
\end{center}
\caption{\small The assembled polarizer. ({\bf a}) Top view, showing the central half-cube shaped body and the crystal and capillary plates holders. The threaded hole is for Helium flowing to reduce the air absorption. ({\bf b}) Side view, with capillary plates in the foreground and the diffracting crystal in the opposite side, hidden from the view. The hole in the middle was made in order to check the alignment between capillary plates and crystal.} \label{fig:Polarizer}
\end{figure}

As stressed in the former section, each component of the source must be easily interchangeable. Hence we placed each crystal and each capillary plate in an aluminum holder. An aluminum pincer was also built for handling the thin capillary plates without damaging the surface. The holders are locked, without allowing for tilt regulation, to the central body of the source which constrains the diffraction angle and is half-cube shaped with 46.5~mm side (see Fig.~\ref{fig:PolarizerTop}). The polarizer, i.e. the block formed by the central body, diffracting crystal and capillary plates which constrain incoming and output radiation, is shown in Fig.~\ref{fig:Polarizer}. Since air absorption heavily affects low energy photons, we reduced to about 5~cm the distance between incoming and output capillary plate. Moreover we plan to employ Helium flowing in the polarizer and then added two standard gas connectors on the top and bottom side of the polarizer, shown in Fig.~\ref{fig:PolarizerTop}. To avoid dust deposition, we protected the capillary plates with 4~$\mu$m thin polypropylene film which assures 96\% transparency at 2.6~keV.

Even if numerically controlled instruments were employed to build every part of the source, we checked the alignment of the capillary plates and diffracting crystal by means of optical measurement with an estimated sensitivity of 5' (about $0.1^\circ$). We locked the polarizer on a manual rotational stage and, rotating at 45$^\circ$ steps, we measured the angle between capillary plates and crystal by means of laser reflection. While the capillary plates resulted aligned within errors, the crystal holder was aligned within about $0.2^\circ$ by means of a thin wedge. The alignment was also verified by measuring the energy of diffracted photons (see Sec.~\ref{sec:Spectrum}).

\begin{figure}[htbp]
\begin{center}
\includegraphics[angle=0, width=9cm]{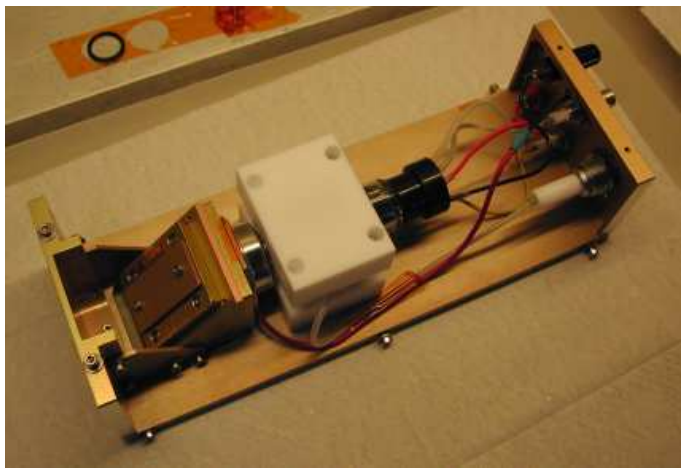}
\end{center}
\caption{\small The complete source of polarized photons. On the left is the polarizer, while the low-power X-ray tube is on the right.}\label{fig:Source}
\end{figure}

The complete polarized source is shown in Fig.~\ref{fig:Source}. The small X-ray tube is locked with a teflon grab to an aluminum flange. The polarizer is independently screwed to the flange and the substitution of capillary plates or crystal doesn't require the disassembling of the X-ray tube. An aluminum case completely covers the source, which is $262\times98\times69~\mbox{mm}^3$.

\subsection{Spectrum and estimated polarization}\label{sec:Spectrum}

\begin{figure}[htbp]
\begin{center}
\includegraphics[angle=0, totalheight=6cm]{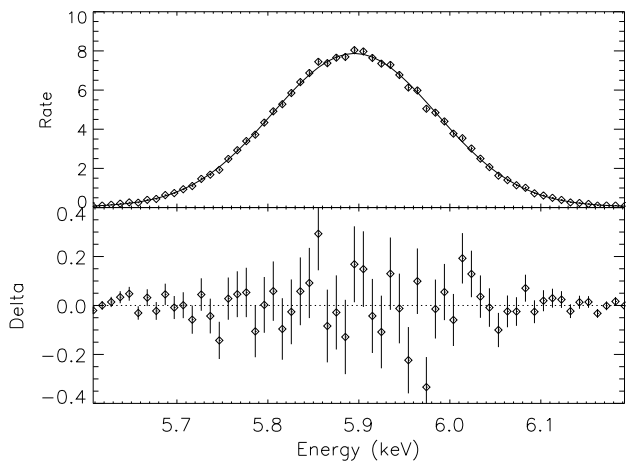}
\end{center}
\caption{\small The spectrum of the 5.9~keV line from a Fe$^{55}$ radioactive source measured with the Amptek XR100CR detector. A gaussian fit provides 213.1~eV FWHM.}\label{fig:FeLine}
\end{figure}

\begin{figure}[htbp]
\begin{center}
\subfigure[\label{fig:GraphiteSpectrum}]{\includegraphics[angle=0,totalheight=5.7cm]{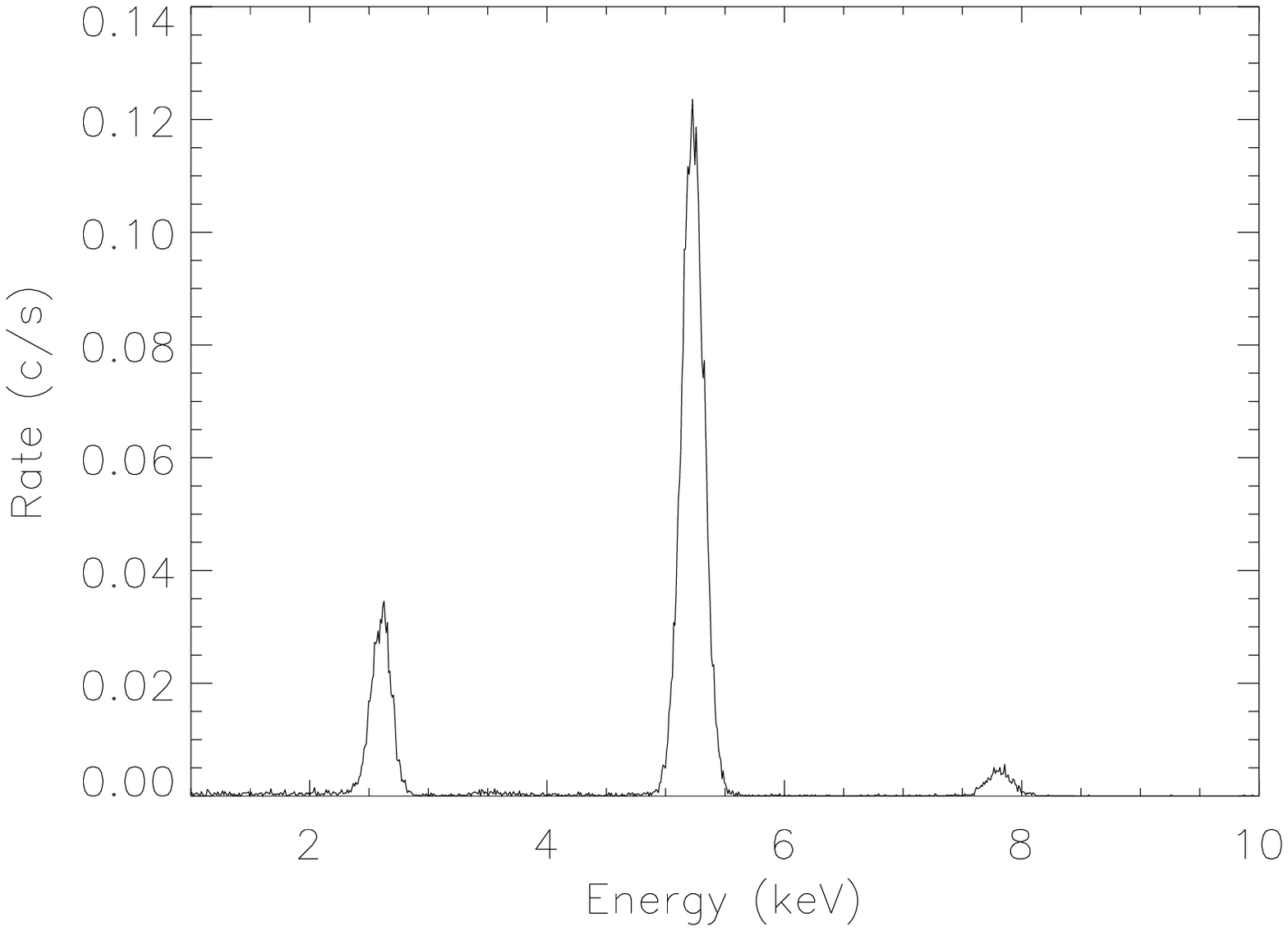}}
\subfigure[\label{fig:Spectrum2cp}]{\includegraphics[angle=0, totalheight=5.7cm]{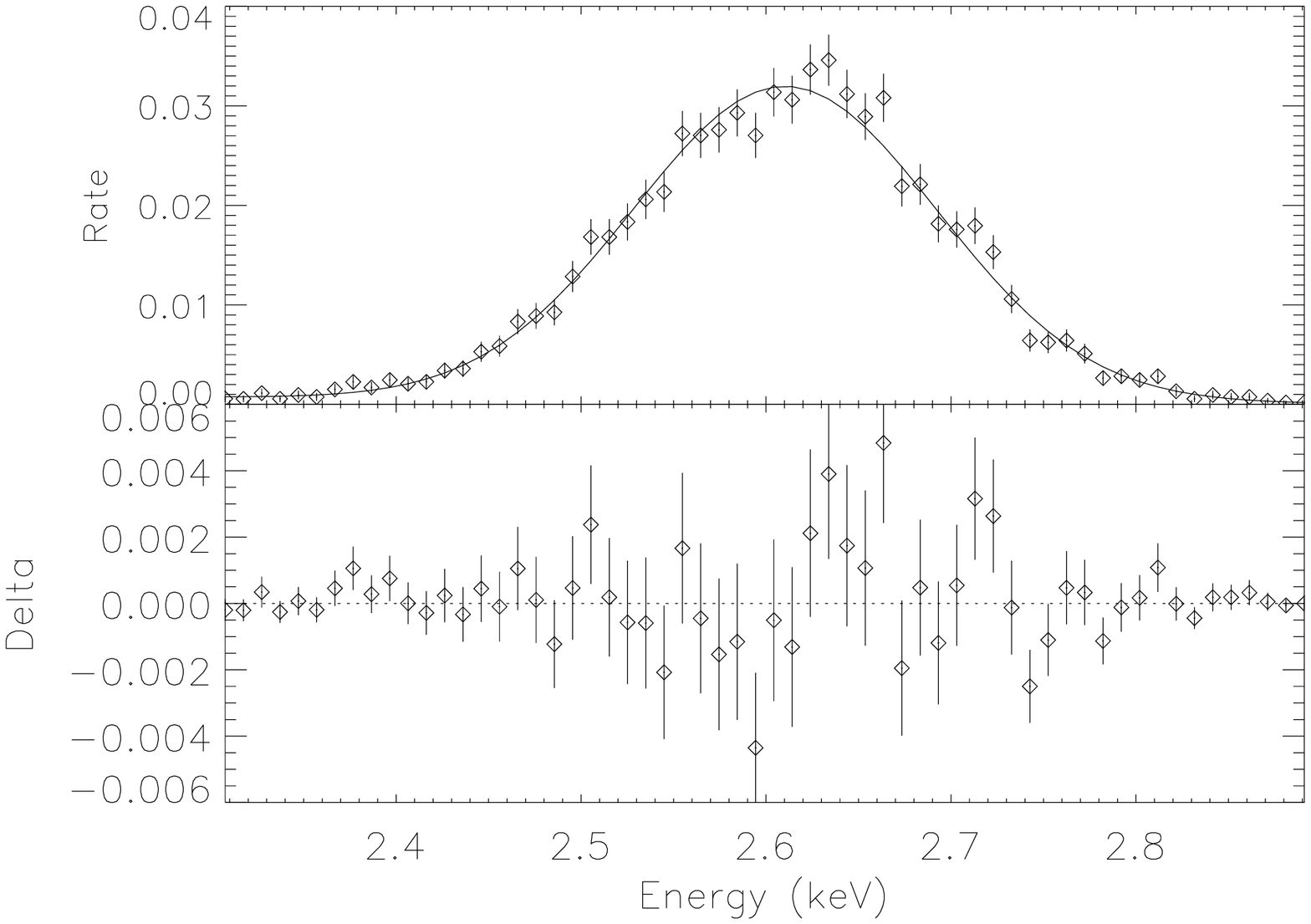}}
\end{center}
\caption{\small The spectrum of diffracted photons from the graphite crystal. Two capillary plates with $\frac{1}{40}$ collimation are employed to constrain incoming and output radiation. ({\bf a}) Spectrum between 1 and 10~keV. Despite the fact that the X-ray tube high voltage is set to 8~kV and hence very low intensity is incident at 7.8~keV, the first three orders of diffraction (2.6, 5.2 and 7.8~keV) are visible with a relative peak height depending mainly on the spectrum of the unpolarized incident radiation and on the air absorption. ({\bf b}) Fit to the first order diffracted photons. See Table~\ref{tab:LineParameters} for fit parameters.}
\end{figure}

Since the energy and the width of the diffracted line can be exploited to estimate its degree of polarization, the spectrum of diffracted photons was measured by means of an Amptek XR100CR Si-PIN detector, with 213~eV resolution at 5.9~keV (see Fig.~\ref{fig:FeLine}). In Fig.~\ref{fig:GraphiteSpectrum} an example of the spectrum of diffracted photons is presented when graphite grade~B and two broad capillary plates (semi-collimation 1.4$^\circ$) are employed. Setting the X-ray tube voltage to 8~kV, continuum photons up the third order (7.8~keV) are diffracted.

The degree of polarization can be estimated from the line energy and Eq.~\ref{eq:PolarizationDegree}. This procedure can be applied only if the line width is very narrow with respect to the variation of $k$, but it can be approximately employed if the line is unresolved from the detector. Otherwise a lower limit to the polarization of the diffracted line can be provided by the line width, since it is a hint of the depolarization occurred due to the diffraction at angles slightly different from 45$^\circ$.

In Fig.~\ref{fig:Spectrum2cp} we report the spectrum of the first order diffracted photons when continuum radiation is incident on a graphite mosaic grade~B crystal. The FWHM of the polarized line is comparable with the detector resolution and the background is negligible and mainly due to the uncompleted charge collection in the silicon detector. The diffraction angle, calculated with Bragg's law from the center of the line, is compatible with 45$^\circ$ within mechanical tolerance (see Table~\ref{tab:LineParameters}).

Removing the capillary plate which constrains the incoming photon direction allows an increase of the counting rate by about a factor of 5. However the reduced collimation and the mosaic structure of the graphite crystal imply the diffraction of continuum incident photons at angles slightly different from 45$^\circ$ and hence an increase of the FWHM and a slight depolarization of the diffracted line. Line fit parameters and polarization estimates, in this case, are reported in the lower part of the Table~\ref{tab:LineParameters}.

The diffraction of calcium K$\alpha$ line on the aluminum crystal is much more effective. Since the contribution of the continuum radiation, measured with the copper X-ray tube, is about a factor 15 less than that of the fluorescence photons, we can assume that incident photons are entirely produced at energy equal to the calcium K$\alpha$ line. Moreover, the flat crystal constrains the diffraction angle much better than the graphite mosaic one, allowing the employment of the outgoing collimator alone without degradation in the polarimetric performance of the source. The lack of contamination from continuum photons can be confirmed from the energy of the diffracted line, which is consistent with the weighted average K$\alpha$1 and K$\alpha$2 calcium lines energy within the 1-$\sigma$ error of about 2~eV. Since the energy of diffracted photons is exactly determined from the calcium K$\alpha$ line energy, the polarization of the diffracted line can be precisely calculated from  Eq.~\ref{eq:PolarizationDegree}, as reported in Table~\ref{tab:LineParameters}.

\begin{table}[htbp]
\begin{center}
\begin{tabular}{|l|c|c|c|c|c|c|c|c} 
\hline
\hline
\multicolumn{6}{l}{Incoming and output $\frac{1}{40}$ collimators} \\
\hline
Diffraction             & Incident radiation & $E$ (keV)                      & FWHM (eV)       & $\chi^2$  & $\theta_{Bragg}$       & $ \mathcal{P}$     & Rate (c/s) \\
\hline
Graphite, I order  & Continuum           & 2.6105$\pm$0.0020       & 193.6$\pm$3.2  & 0.849         & 45.07$^\circ$            & $\sim0.98$           & 0.66\\
Graphite, II order & Continuum           & 5.2261$\pm$0.0018       & 220.1$\pm$1.6   & 1.101        & 45.02$^\circ$            & $>0.96$                & 2.8\\
\hline
\hline
\multicolumn{6}{l}{Output $\frac{1}{40}$  collimator} \\
\hline
Graphite, I order & Continuum            & 2.6109$\pm$0.0036        & 198.6$\pm$4.7   & 1.096       & 45.07$^\circ$           & $>0.95$                & 3.5\\
Graphite, II order & Continuum           & 5.2269$\pm$0.0037       & 248.2$\pm$2.6    & 0.926       & 45.01$^\circ$            & $>0.94$                & 16.8\\
Aluminum, I order & calcium K$\alpha$ line & 3.6889$\pm$0.0024 & 200.1$\pm$0.2  & 1.396      & 45.93$^\circ$           & $0.9938$              & 140.1\\
\hline
\end{tabular}
\caption{Fit parameters, diffraction angle and estimated polarization of the diffracted lines for different collimators and crystals configurations. The diffraction angle for graphite grade~B crystal and hence continuum radiation, calculated from peak energy and Bragg's law, is compatible within mechanical tolerance with 45$^\circ$ diffraction. The increase of the FWHM with energy is partially due to the decreasing resolution of the detector. Air absorption is not taken into account.} \label{tab:LineParameters}
\end{center}
\end{table}

\section{The modulation factor of the Gas Pixel Detector at low energy} \label{sec:Result}

The polarized source described in Sec.~\ref{sec:BraggSource} was employed to measure the modulation factor of the 105k pixels, 50~$\mu$m pitch, sealed version of the Gas Pixel Detector, described by \cite{Bellazzini2006c} and filled with 20\% Helium and 80\% DME mixture. Data were analyzed in the standard way, without any optimization for the reconstruction of low energy photoelectrons tracks.

Since the estimated degree of polarization is nonetheless very high, all the measurements were performed with the outgoing collimator alone to decrease the integration time, placing the source as close as possible to the detector to reduced the air absorption. The measurements at 2.6 and 3.7~keV were performed setting the X-ray tube high voltage below the energy of the second order of diffraction to avoid contamination of higher order diffracted photons. In the analysis at 5.2~keV, first order photons were removed from off-line selection of the energy of the event.

The spot of diffracted photons on the $15\times15~\mbox{mm}^2$ imaging Gas Pixel Detector was limited by the diameter of the incident beam to about 5~mm. Only events in this region were analyzed. Standard cuts were applied to discard the less asymmetric photoelectrons tracks which would be hardly reconstructed. Selective cuts increase the modulation factor, but reduce the efficiency of the detector. Recalling that the minimum detectable polarization is inversely correlated with $\mu\sqrt{\epsilon}$, a careful data selection is required to maximize the MDP.

\begin{figure}[htbp]
\begin{center}
\subfigure[]{\includegraphics[angle=0,height=4.5cm]{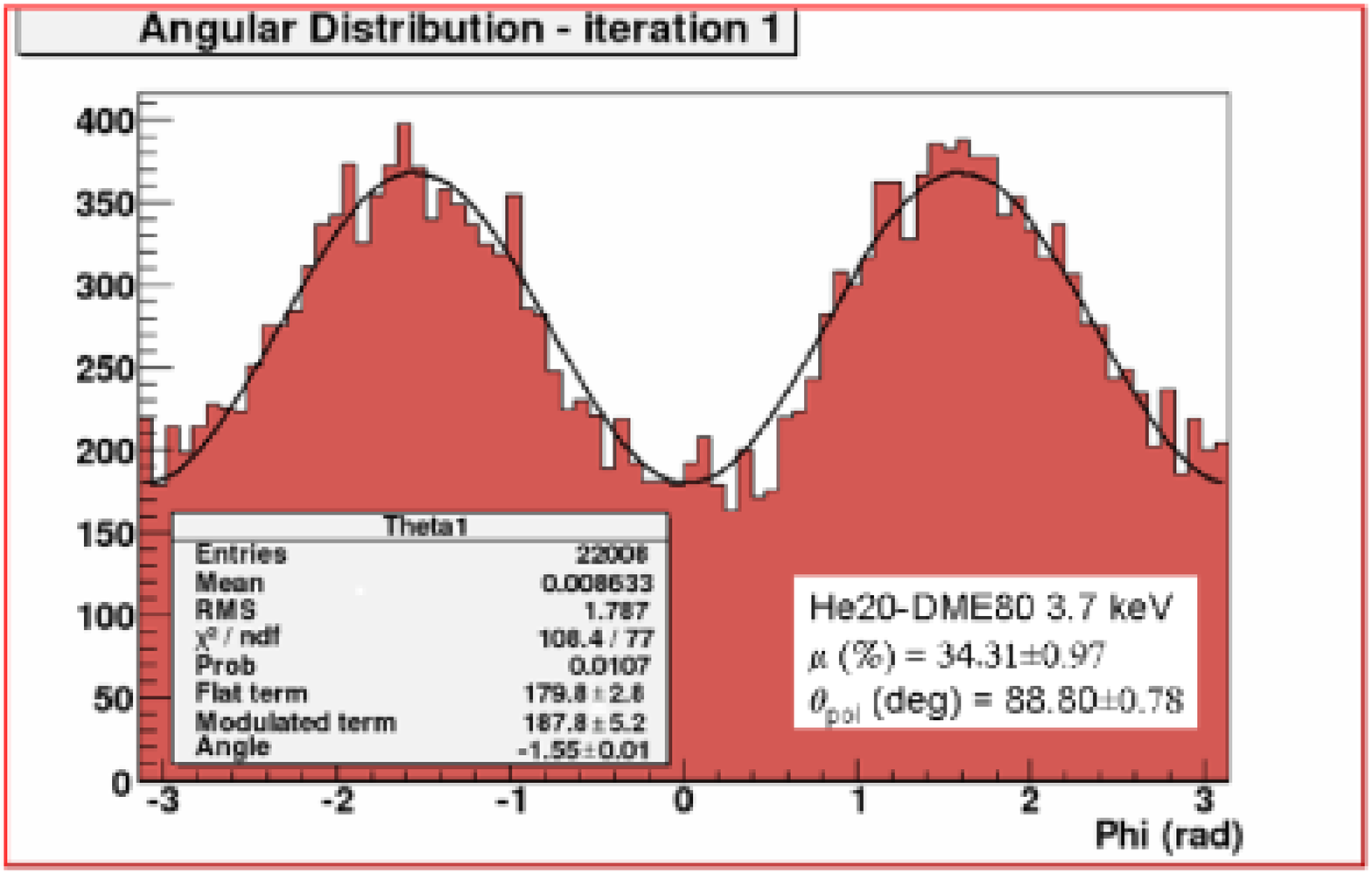}}
\subfigure[]{\includegraphics[angle=0, height=4.5cm]{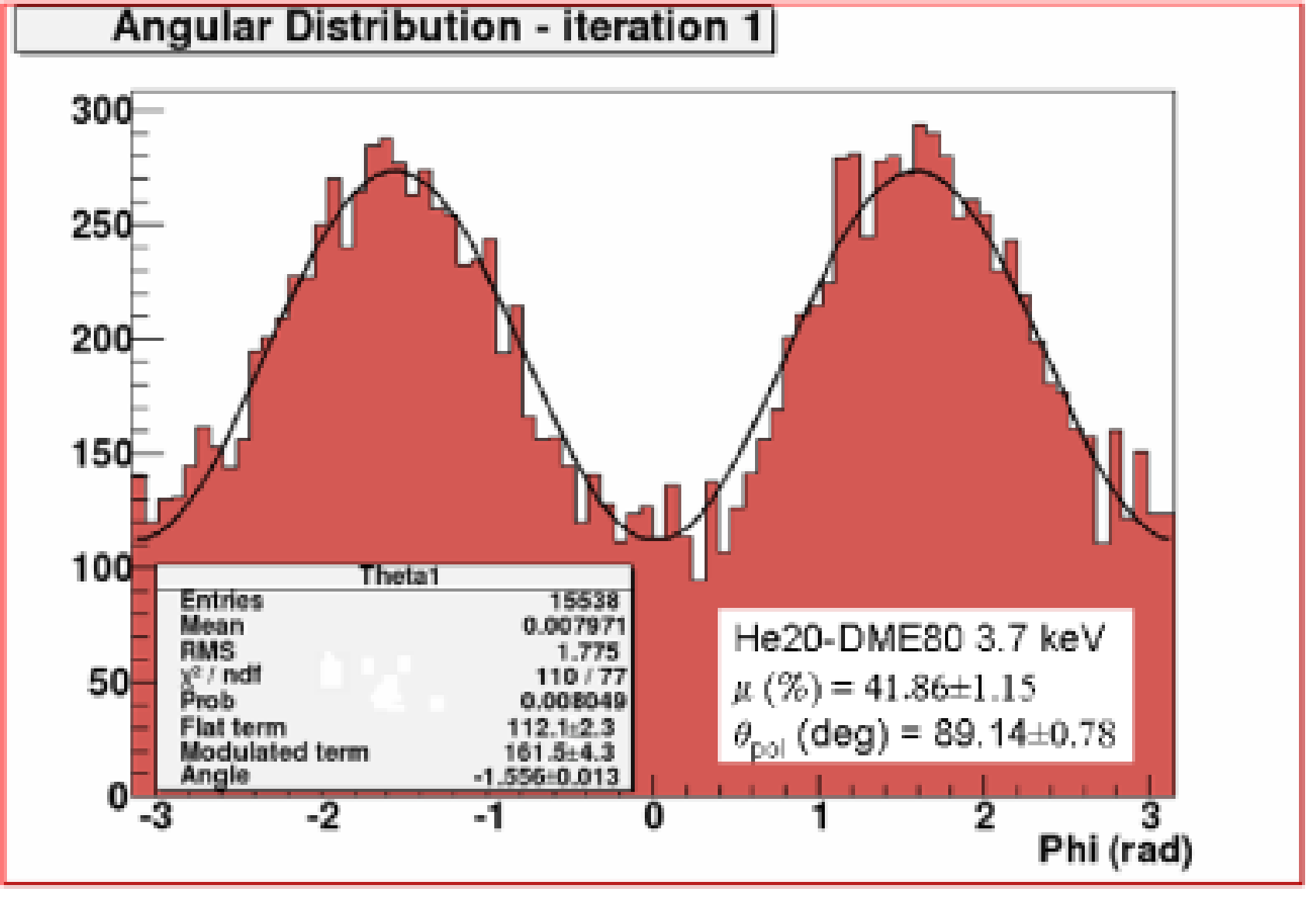}}
\end{center}
\caption{\small Increase of the modulation factor with data cuts. Modulation of reconstructed angles at 3.7~keV without any cuts  ({\bf a}) and with cuts implying a reduction of the efficiency at 71\%  ({\bf b}). \label{fig:Modulation}}
\end{figure}

The modulation factor measured at 2.6, 3.7 and 5.2~keV with and without data cuts (see fig.~\ref{fig:Modulation}) is reported in Table~\ref{tab:Modulation}. The relative reduction of the efficiency $\epsilon_R$ due to the cuts is also reported. Polarization angle was changed to check possible systematic errors due to the hexagonal pattern of the pixels. The values at 2.6 and 5.2~keV are corrected assuming that incident radiation is 0.95 polarized (see the lower part of the Table~\ref{tab:LineParameters}).


\begin{table}[htbp]
\begin{center}
\begin{tabular}{c|c|c|c|c}
E (keV)                              & PA                               & $\mu$                             & $\epsilon_R$        & $\mu\sqrt{\epsilon}$ \\
\hline
\multirow{4}{*}{2.6}           & $\sim10^\circ$          & $0.2106\pm0.0119$       & 1                              & 0.0927 \\
                                           & $\sim10^\circ$           & $0.2205\pm0.0122$       & 0.96                        & 0.0951 \\
                                           & $\sim10^\circ$           & $0.2614\pm0.0138$        & 0.77                       & 0.1010 \\
                                           & $\sim10^\circ$           & $0.2762\pm0.0143$        & 0.69                        & 0.1010 \\
\hline
\multirow{3}{*}{3.7}          & $\sim0^\circ$             & $0.3431\pm0.0097$        & 1                             & 0.1039 \\
                                          & $\sim0^\circ$              & $0.4186\pm0.0115$       & 0.71                        & 0.1068\\
                                          & $\sim0^\circ$             & $0.4310\pm0.0118$         & 0.67                        & 0.1068 \\
\hline
\multirow{2}{*}{5.2}           & $\sim0^\circ$            & $0.4653\pm0.0095$        & 1                             & 0.0884 \\
                                           & $\sim0^\circ$            & $0.5447\pm0.0104$        & 0.78                        & 0.0914 \\
\end{tabular}
\caption{Measured modulation factor of the Gas Pixel Detector at different energies. Data cuts, implying an increase of the modulation factor and an efficiency reduction of a factor $\epsilon_R$, were varied to maximize the quality factor, reported in the last column. PA is the approximate value of the polarization angle, changed to check possible systematic effects. The values of the modulation factor at 2.6 and 5.2~keV assume that incident polarization is 0.95 polarized.} \label{tab:Modulation}
\end{center}
\end{table}

\begin{figure}[htbp]
\begin{center}
\includegraphics[angle=0, totalheight=6cm]{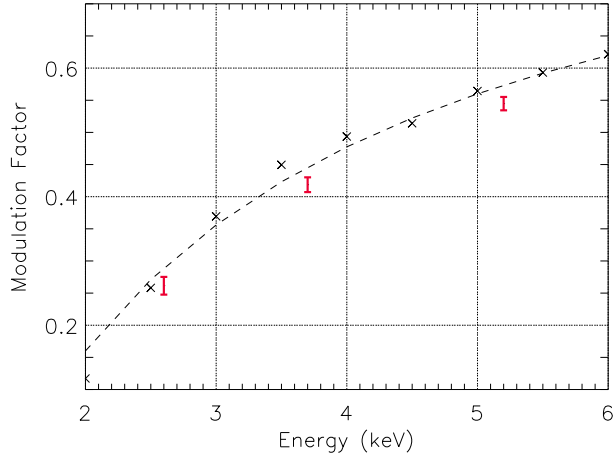}
\end{center}
\caption{\small Comparison between modulation factor expected from the Monte Carlo simulations (crosses) and the measured values at 2.6, 3.7 and 5.2~keV (red). It is assumed that the degree of polarization at 2.6 and 5.2~keV is 0.95. Data selection, with an efficiency reduction of 0.77, 0.71 and 0.78 at 2.6, 3.7 and 5.2~keV, is also applied. The dashed line is the fit to Monte Carlo values.}\label{fig:ModulationFactor}
\end{figure}

In Fig.~\ref{fig:ModulationFactor} the comparison between the expected modulation factor, derived from the Monte Carlo software, and the measured values is reported. Cuts on the measured data, with an efficiency reduction at 0.77, 0.71 and 0.78 at 2.6, 3.7 and 5.2~keV respectively, were applied. The agreement is very good, validating the Monte Carlo software employed till now and proving that the Gas Pixel Detector is sensitive to polarized radiation even at low energy, i.e. in the range of maximum interest. These measurements also confirm the fast increase of the modulation factor with energy.
It is important to note that the agreement between expected and measured modulation factor for the He-DME gas mixture gives us much more confidence with the Monte Carlo results. This was of particular importance, since the Monte Carlo software is intensely used for the development of the detector and for the choice of the best mixture in a selected energy range.

Despite this good agreement, we can observe that the measured modulation factor results slightly lower than that predicted. Even if a further investigation is needed, this could be due to the 1.2~mm gap between the Gas Electron Multiplier and the collection plane, which is not yet fully implemented in the Monte Carlo software. A new version of the detector with this distance reduced to 400 $\mu$m has been recently built and will soon be tested.

\section{Conclusion} \label{sec:Conclusion}

We presented a new polarized source, based on the Bragg diffraction at 45$^\circ$, built for the study of the polarimetric sensitivity of the Gas Pixel Detector. Unpolarized photons are diffracted on a crystal and, constraining the diffraction angle at nearly 45$^\circ$, only the component polarized orthogonally to the plane of incidence is efficiently diffracted. The employment of capillary plates, allow, despite their very limited size, an effective constrain on the diffraction angle. The output radiation is almost completely polarized and also nearly monochromatic, since the diffraction angle and photon energy are related by the Bragg's law within some eV. A suitable choice of the diffracting crystal enables the production of polarized X-ray photons even at very low energy. Using graphite and aluminum crystals for diffraction, we produced nearly completely polarized photons at 2.6, 3.7 and 5.2 keV, with an energy line width of some tens of eV.

This effective and low energy polarized source was employed to measure the polarimetric sensitivity of the Gas Pixel Detector between 2.6 and 5.2~keV, where the maximum performances are expected. The measured modulation factor is in good agreement with the Monte Carlo results. This not only validates the tools and the estimates of the modulation factor made so far with the Monte Carlo software, but also confirms that the X-ray photoelectric polarimetry is now feasible.

No low energy optimization of the data analysis, which was tuned on longer photoelectron tracks, was performed yet. However the results are rather encouraging. Currently we are able to reach the $\sim$80\% of the polarimetric sensitivity, i.e. the quality factor, expected from the Monte Carlo estimates. Moreover, for the first time, we now have an effective low energy polarized source that can be exploited to achieve a further improvement of the performance of the detector in the region of maximum sensitivity.

\section*{Acknowledgments} 

This research is supported by ASI contract I/074/06/0 and INAF PRIN 2005.

\bibliography{References}
\bibliographystyle{unsrt}

\end{document}